 \documentstyle[twocolumn,aps,epsf]{revtex}
 \begin{document}
 \title{Ballistic spin transport through electronic stub tuners: \\
 spin precession, selection,  and square-wave  transmission}

 \author{ X. F. Wang$^\star$, P. Vasilopoulos$^\star$, 
 and F. M. Peeters$^\diamond$ 
 \ \\ }
 \address{$^\star$Concordia University, Department of Physics,
 Montr\'{e}al,  H3G  1M8, Canada\\
 \ \\
 $^\diamond$Departement Natuurkunde, Universiteit Antwerpen (UIA),\\
Universiteitsplein 1, B-2610, Belgium }
 \address{}
 \address{\mbox{}}
 \address{\parbox{14cm}{\rm \mbox{}\mbox{}\mbox{}
Ballistic spin transport is studied through electronic tuners with double stubs
attached to them. The spins precess due to the spin-orbit interaction. Injected
polarized spins can exit the structure polarized in the opposite direction. A
nearly square-wave spin transmission, with values 1 and 0, can be obtained
using a periodic system of symmetric stubs and changing their length or width.
The gaps in the transmission can be widened using {\it asymmetric} stubs. An
additional modulation is obtained upon combining stub structures with
 different values of the spin-orbit strength.\\}}
  \address{\mbox{}}
 \address{\parbox{14cm}{ \rm PACS 72.20.-i, 72.30.+q, 73.20.Mf}}
  \maketitle

The basic principle of a spin transistor, first  formulated in Ref. \cite{das}
for a waveguide, is that  the spin-orbit interaction or Rashba coupling
\cite{ras}, important in narrow-gap semiconductors, makes injected polarized
spins precess and leads to a modulation of the current. It's been demonstrated
that the strength of this interaction can be tuned by the application of an
external gate voltage \cite{nit}. In recent years spin polarized transport has
attracted considerable attention  as it offers a possibility for quantum
computation and quantum logic  \cite{tod}. However, the reported
experimental spin polarizations \cite{ham} are very low, about 1\%, and make
the results controversial since they can be attributed to extraneous effects
such as the local Hall field and the resistance mismatch \cite{tan}.

The idea of ballistic spin transport of Ref. \cite{das} relied on the weakness
of the spin-orbit coupling. 
It was  recently applied to nanowires by means of a tight-binding analog of the
Rashba Hamiltonian \cite{mir} thought to be an improvement over a perturbative
treatment. A perfect 
modulation was reported for {\it weak} coupling.
However, the results were tied to a gradual change of the coupling strength
over the interaction region, which may be difficult to achieve experimentally,
and those for {\it strong} coupling may be uncertain due to the large 
strengths
used.

The subject of this paper is a transistor-like modulation of a spin
current. Motivated by the results of Refs.
\cite{das} and \cite{mir} and those 
on electronic \cite{deb},
 \cite{tak1}, \cite{pet}, and
photonic \cite{tak2} stub tuners, we consider ballistic spin transport through
electronic waveguides  with double stubs attached to them ( Fig. 1 (a)) periodically. The weakness
of the spin-orbit coupling is controlled by  back gates \cite{nit}. 

{\it Formulation}. In the absence of a magnetic field the spin degeneracy of the 
2DEG  energy bands at ${\bf k}\ne 0$ is lifted by the
 coupling of the electron spin with its orbital motion. This coupling is
 described by the Hamiltonian

 \begin{equation}
 H_{so}=\alpha (\vec{ \sigma }\times
 \vec p)_z/ \hbar=i\alpha [\sigma _y \partial
 /\partial x-\sigma _x \partial/\partial y)].
 \end{equation}
 Here the $y$ axis is along the waveguide and the $x$ along the stub, cf.
Fig. 1 (a).  The parameter $\alpha$  measures the
 \begin{figure}[t]
 \begin{center}
 \addvspace{0 cm}
 \leavevmode
 \epsfysize=125pt
 \epsfxsize=235pt
 \epsffile{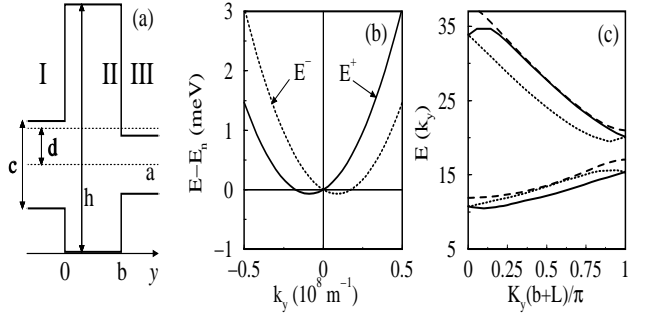}
 \end{center}
 \caption{
 \label{f1} (a) Schematics of the stub tuner. (b) Dispersion relation for a
 waveguide based on Eqs. (4) and (5). (c) Superlattice dispersion relation
 with $\alpha$ equal to  zero for the dashed curves and finite
 for the solid ($E^+$) and dotted ($E^-$) curves.}
 \end{figure}
\noindent
 strength of the coupling;  
 $\vec{\sigma}=(\sigma_x,\sigma_y,\sigma_z)$  denotes
 the spin Pauli matrices,  and $ {\vec p}$ is the momentum
 operator. The experimental values of $\alpha$ 
range from about $6\times 10^{-12}$ eV\,m,  at electron densities
 $n=0.7\times 10^{12}\, $cm$^{-2}$,  to $3.0 \times 10^{-11}$\,eV\,m,
 at  electron densities $n=2\times 10^ {12}\,$cm$^{-2}$.

 We treat  $H_{so}$ as a perturbation. With
 $\Psi =| k_{y},n,\sigma\rangle
 =e^{ik_{y}y}\phi_{n}(x)|\sigma\rangle$  the
 eigenstate in each region in Fig. 1 (a) the
 unperturbed 
 states satisfy 
 $H^{0}|n,\sigma\rangle=E_{n}^{0}|n,\sigma\rangle$ with
 $E_{n}^{0}=E_{n}+\hbar^2k_{y}^{2}/2m^*$ and 
 $\phi_{n}(x)$ 
 obeys
$[-(\hbar^2/2m^*) d^2/dx^2+V(x)]\phi_n(x)=E_n\phi_n(x)$,
 where $V(x)$ is the confining potential assumed to be square-type and
high enough that  $\phi_n(x)$ vanishes at the boundaries.
 The perturbed ($H_{so}\neq 0)$ eigenfunction,
 is written as $\sum_{n,\sigma}A^{\sigma}_n\phi_n(x)|\sigma\rangle$. $H_{so}$ is
a $2\times 2$ matrix. Combining it with  
 the  $2\times 2$ diagonal
 matrix $H^{0}$ and using  $H\Psi=(H^{0}+H_{so})\Psi=E\Psi$ leads to the equation

 \begin{equation}
 \left [ \begin{array}{cc} E_{m}^{0}-E&
  \alpha k_{y} \\ \alpha k_{y} & E_{m}^{0}-E
 \end{array} \right ]\left
 ( \begin{array}{c} A^+_m \\
 A^-_m \end{array} \right )=0; 
 \end{equation}
 the resulting eigenvalues $E\equiv E^{\pm}(k_y)$, plotted in Fig. 1 
 (b), are 

 \begin{equation}
 E^{\pm}(k_y)=E_n+(\hbar^2/2m^*)k_y^2\pm\alpha k_y.
 \label{disp}
 \end{equation}
 The eigenvectors corresponding to $ E^{+}, E^{-}$ satisfy $A_{m}^\pm
 =\pm A^\mp_{m}$. Accordingly, the spin eigenfunctions are taken as
$|\pm\rangle={\tiny\left(\array{c} 1 \\ \pm 1 \endarray \right)}/\sqrt{2}$.
For the same energy the difference in wave vectors $k^+_{y}$ and $k^-_{y}$ for
the two spin orientations is

 \begin{equation}
 k^-_{y}-k^+_{y}=2m^*\alpha/\hbar^2= \delta.
 \label{difk}
 \end{equation}
The dispersion relation $E^\pm(k_y)$ vs $k_y$ resulting from Eq.
(\ref{disp})
is shown in Fig. 1 (b). For the same energy $E$ there are four $k_y$ values
and a phase shift $\delta$ between the positive or negative $k^+_y$ and
$k^-_y$ values  
of the branches $E^+$ and $E^-$.

 The procedure outlined above applies to all regions,
 I, II, III , in Fig. 1 (a). In each region we have
 $\phi_{n}(x)=\sin(n\pi (x+w/2)/w)$, where
 $w$ is the width of the region along $x$. Including spin and referring to
 Fig. 1 (b) we can write the eigenfunction $\phi_1$ of energy $E$
 in region I as

\begin{eqnarray}
\nonumber
&&\phi_1 =\sum_m \{ a^+_{1m}e^{i\beta_my}( \begin{array}{c} 1\\  1
\end{array})
+a^-_{1m}e^{i(\beta_m+\delta)y}( \begin{array}{c} 1\\  -1 \end{array})
+b^+_{1m}\\
&\times& 
e^{-i(\beta_m+\delta)y}  ( \begin{array}{c} 1\\
1 \end{array})
+b^-_{1m}e^{-i\beta_my}(\begin{array}{c} 1\\  -1 \end{array})\} 
\sin  (c_m (x+\frac{c}{2})
\label{wavf}
\end{eqnarray}
Here $c_m=m\pi/c$ and $\beta_m=(2m^*E-c_m^2)^{1/2}$.
In region III $\phi_2$ is given by Eq. (\ref{wavf}) 
with the changes $1m\to 2m, c\to a$, and $y\to y-b$. In the stub region II,
Eq. (\ref{wavf}) remains valid with the changes
$c\to h$ and $x+c/2\to x+h/2-d$.

 We now match the wave
 function and its derivative at $y=0$ and $y=b$. In this way we can
 connect the incident waves (to the left of region I) with the
 outgoing ones (to the right of region III) via a transfer matrix $\hat{M}$

 \begin{equation}
 \left ( \begin{array}{c} a^+_{in}\\ a^-_{in}\\ b^+_{in}\\
 b^-_{in} \end{array} \right )
 =\hat{M} \left ( \begin{array}{c} a^+_{out}\\ a^-_{out}\\
b^+_{out}\\
 b^-_{out} \end{array}
 \right ).
 \end{equation}

 If we connect a spin polarizer (analyzer) to the left (right) of the
structure, we can inject 
electrons 
and detect the  polarization of the outgoing electrons.
For spin-up electrons injected into a simple waveguide, where the transmission
is always unity, the
probability of detecting a spin-down 
${\tiny \left( \array{c} 0 \\ 1 \endarray\right)}$ electron, after
a distance $l$,
 will be proportional to 
$|\langle \tiny \left( \array{c} 0\\ 1\endarray \right)|\psi\rangle|^2
=\sin^2(\Delta\theta/2)$ \cite{das,mir}, where $\Delta\theta=\delta l$ is the phase
difference between the up $|+\rangle$ and down $|-\rangle$ spin  modes
Here we show
that by attaching stubs the transmission can be  modulated more efficiently: 
we can  flip the spin
of the incident electrons, or block it completely,
and thus establish a spin transistor.

{\it Results}.   We consider electrons of energy $E=48$meV injected into a
In$_{0.53}$Ga$_{0.47}$As  multi-stub structure, with 
$m^*=0.042m_0$ and 
$\alpha=1.6\times10^{-11}$eVm.
The parameters are $c=a=250$\AA, $b=150$\AA, and $h=1859$\AA;\
 the length of the waveguide segment between two neighbouring stubs is
$L=207.5$\AA. To verify the validity of the
perturbation theory  we evaluated the bound states of an isolated unit made
of one stub and one  waveguide segment with the same values for $c, a, b$ and
$L$. 
The ratio 
(intersubband mixing
energy/
difference between  lowest two bound states) is less than
$10\%$. In  Fig. 1 (c) we show 
 the first and  second energy bands
of a superlattice made of such units
without (dashed curves) and with (solid
and dotted curves) spin-orbit coupling.
The energy difference due to the latter is about $15\%$ of the electron energy.
 \begin{figure}[t]
 \begin{center}
 \addvspace{0 cm}
 \leavevmode
 \epsfysize=118pt
 \epsfxsize=200pt
 \epsffile{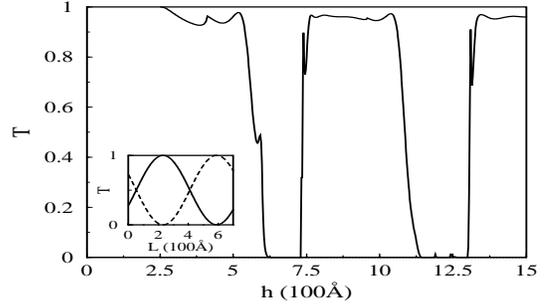}
 \end{center}
 \caption{
 \label{f2} Transmission $T$ vs stub length $h$. The inset shows $T^{-}$
(solid curve) and $T^{-}$  
(dotted curve)
electrons, normalized to the input flux  $T^{+}$, 
vs the ({\it
stubless}) waveguide's length $h$.}
 \end{figure}
We now consider  the possibility of spin-transistor action using {\it
symmetric} stubs. The transmission plotted in Fig. 2  shows the possibility of
spin blocking as well as spin flipping for a  periodic array of fifteen stubs
arranged in   three groups. The first
five stubs form the first group and so on. The  parameters $c=a=250$\AA,
$L=267.5$\AA, \
and $h$\
 are the same for all groups but 
 $\alpha$ and $b$ differ from
one group to another: we took $\alpha_1=1.05\alpha_2=1.1\alpha_3=1.6\times
10^{-11}$eVm  and $b_1=0.95b_2=0.9b_3=150$\AA. Because the output spin
orientation depends on the total length of the device, we choose the length
for which complete spin flip occurs through the {\it stubless} device; then
we can  control
the transmission by adjusting $h$. As shown in the figure,
we can completely block the exit of electrons  of either spin orientation
for $h$ in the ranges of the gaps ($T=0$), i.e., for
$620\AA \leq h\leq 680\AA$   and $1130\AA \leq h\leq 1285\AA$.\\
 \hspace*{.5 cm}A  better control of the transmission can be obtained if we employ {\it
asymmetric} stubs. For the results shown in Fig. 3 we inject spin-up electrons of energy 
$E=48meV$ into a five-identical-unit  device with $c=a=250$\AA, 
 \begin{figure}[t]
 \begin{center}
 \addvspace{0 cm}
 \leavevmode
 \epsfysize=118pt
 \epsfxsize=200pt
 \epsffile{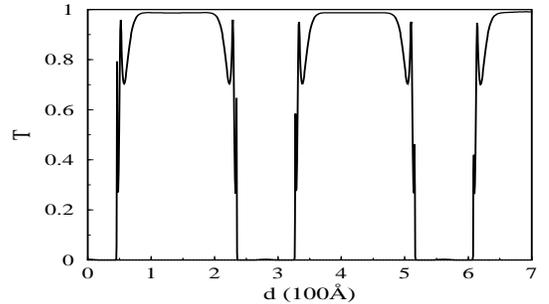}
 \end{center}
 \caption{ \label{f3} Transmission $T$ vs asymmetry
 parameter $d$. }
 \end{figure}
\noindent   
 $b=150$\AA, $L=207.5$\AA, \   
and      $h=1859$\AA \
for all units.  $L$  
is chosen so that only spin-down electrons appear in the output and 
$h$ 
is chosen such that all electrons are reflected ($T=0$) when the stubs
are symmetric.
As realized, by shifting all stubs by a distance $d$, cf. Fig. 1 (a),
%
with the help of side gates, we  observe a nearly perfect spin-transistor
behavior: the transmission 
jumps almost from 0 to 1 and only
spin-down electrons come out 
though only spin-up electrons are injected. 

{\it Discussion}. 
A qualitative understanding of the results shown in  Figs. 2 and 3 
is easily reached if we combine
the spin precession in a single waveguide \cite{das}, due to the spin-orbit coupling,
with the basic idea of a stub tuner \cite{hes} and its refinements 
\cite{tak1}. In a stub tuner waves reflected from the walls of the stub,
where the wave function vanishes, may interfere constructively or destructively
with those in the main waveguide and result, respectively, in an increase  or
decrease of the transmission. Refining this idea,  it was shown in Ref.
\cite{tak1} that using {\it asymmetric} double stubs the transmission of {\it spinless}
electrons could be blocked completely. 
Combining several 
stubs would result in a nearly square-wave transmission as a
function of the asymmetry parameter $d$. The transmission shown in  Figs. 2 and
 3 is simply the result of  this behavior  when combined with the spin
precession due to the spin-orbit coupling since the length of the device
was  chosen such that spin flip would occur in the stubless waveguide.

An important question is how robust  the results are if we change any of the
stub parameters. As shown in Figs. 2 and 3 the 
transmission is not always perfect: near the edges of the gaps  we have 
peaks less high
than unity and their number increases if we change, 
e.g., $h$
in Fig. 2 or $d$ in Fig. 3. However, 
the gaps are wide and one can  widen them further 
by adjusting 
$b$, $c$, and especially
the strength $\alpha$ that can be controlled by a back gate.

Another question is the influence of the stub shape  on  
the transmission output. But as in 
electronic stub
tuners \cite{tak1} , here two stubs of  different shape
do not change the transmission qualitatively.  We break each shape in 
a 
\begin{figure}[t]  
\begin{center}  
\addvspace{0 cm}
 \leavevmode
 \epsfysize=125pt
 \epsfxsize=200pt
 \epsffile{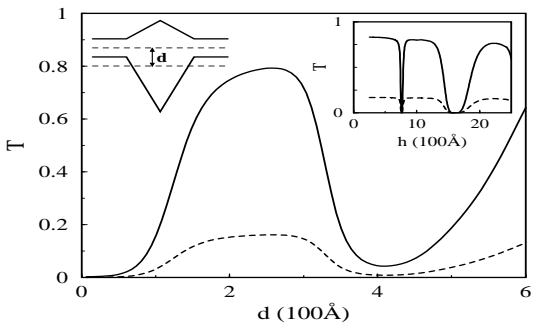}
 \end{center}
\caption{ \label{f4} Transmission $T$ through two {\it asymmetric}, triangular
double stubs 
vs asymmetry parameter $d$. The solid (dotted)
curve is for spin-down (spin-up) electrons. 
The right inset shows $T$ vs $h$ 
through two {\it symmetric},
triangular  stubs.}
\end{figure}
\noindent   series of rectangular segments 
with the same width $b_i$ and different heights. Each
segment  is described by a transfer matrix $M_i$. The full shape is
well described by the product $M^{tot}=\Pi_iM_i$ if $b_i$  is much
smaller than the electronic wavelength. In Fig. 4 we consider a structure with
two {\it asymmetric} triangular double stubs, as shown in the left inset, and 
plot $T$ vs $d$, 
normalized to the input
spin-up $T^{+}$  and further specified in the caption, 
for $b=377$\AA,   $h=1660$\AA,  $L=275$\AA, and $a=250$\AA. 
The total length
$l$ ($=2b+2L$) 
is chosen such that  neither
$T^{+}$   nor $T^{-}$    is completely suppressed.
If $l$ or $\alpha$  is chosen such that, e. g., $T^{+}$  is completely
suppressed, then  $T^{-}$   is given
approximately by the  sum of
the solid and dotted curves in Fig. 4 and resembles closely  that
of Fig. 3.
By comparing Fig. 2 with the right  inset  of Fig. 4 and Fig. 3 with Fig. 4
we see that the qualitative behavior for stubs of different shapes
is the same. The shape affects mainly the period 
of the transmission when we combine several stubs, compare Figs. 2-4.

	All the results presented so far are valid when 
only a single mode propagates
in the waveguide. If more  modes are allowed to propagate
 the transmission pattern
becomes more complex but it is still possible to have a periodic
transmission output, e.g., as in Fig. 2, if $b$ is short enough
that only a single  mode can penetrate into the stub region \cite{tak1}.
Details will be given elsewhere.

        In summary, we  combined the spin precession in a waveguide, due to
the spin-orbit coupling, with the basic physics of a stub tuner, and applied
it to the transmission through several stubs. 
We showed that 
we can select the spin of the outgoing electrons to be the same
as or opposite to that of the injected spin-polarized electrons. 
More important,  we can have a
nearly binary square-wave transmission (spin-valve
effect) for either spin orientation.
In this respect, {\it asymmetric}
stubs, with shape controlled by lateral gates \cite{deb}, give the best results.

 X. F. W. and  P. V. were supported by the  Canadian NSERC Grant No.
OGP0121756  and 
F. M. P. by 
the Flemish Science Foundation, 
 IMEC, IUAP-IV, and by GOA.



\begin{references}


 \bibitem{das} S. Datta and B. Das, Appl. Phys. Lett. {\bf 56},
 665, (1990).

 \bibitem{ras} E. I. Rashba,  Sov. Phys. Solid State {\bf 2}, 1109 (1960)].

 \bibitem{nit} J. Nitta {\it et al.},
 Phys. Rev. Lett. {\bf 78}, 1335 (1997).

 \bibitem{tod} {\it Physics Today}, April 2000, p. 21;
 J. M. Kikkawa and D. D. Awschalom,  Phys. Rev. Lett. {\bf 80},
 4313 (1998);

  \bibitem{ham} P. R. Hammar {\it et al.},
  Phys. Rev. Lett. {\bf 83}, 203 (1999);\\
  S. Gardelis {\it et al.},
 Phys. Rev. B {\bf 60}, 7764 (1999).

 \bibitem{tan} H. X.  Tang {\it et al.},
 Phys. Rev. B {\bf 61}, 4437 (2000); \\G. Schmidt {\it et al.},
 {\it ibid} {\bf 62}, R4790 (2000).

 \bibitem{mir} F. Mireles {\it et al.}, 
 Phys. Rev. B {\bf 64}, 024426
(2001). 

\bibitem{deb} P. Debray {\it et al.},
Appl. Phys. Lett. {\bf 66}, 3137 (1995).

 \bibitem{tak1} R. Akis {\it et al.}, 
 Phys. Rev. B {\bf 52}, 2805 (1995).

 \bibitem{pet} F. M. Peeters, in {\it Science and Engineering 
 of 1- and
 0-dimensional Semiconductors}, Eds. S. P. Beaumont, C. M.
 Sotomayor-Tores (Plenum Press, N. Y., 1990), p. 107.

 \bibitem{tak2} R. Akis {\it et al.}, 
 Phys. Rev.  E {\bf
 53}, 5369 (1996).

\bibitem{hes} F. Sols {\it et al.}, 
 Appl. Phys. Lett. {\bf 54}, 350 (1989).
 \end{references}
\end{document}